\documentclass[twocolumn,showpacs,preprintnumbers,amssymb]{revtex4}
\usepackage{graphicx}
\usepackage{dcolumn}
\usepackage{bm}

\begin{document}
\preprint{PRL/LF8490}

\title{A New Source of Reaction - Diffusion 
Coupling in Confined Systems due to Temperature Inhomogeneity} 

\author{ A. V. Anil Kumar$^{1}$, S. Yashonath$^{1,2}$ 
and  G. Ananthakrishna$^{3,2}$}
                            
\affiliation{$^{1}$ Solid state and Structural Chemistry Unit,
$^{2}$ Center for Condensed Matter Theory,
$^{3}$ Materials Research Center,
Indian Institute of Science, Bangalore, India - 560 012}
\date{\today}
\begin{abstract}
Diffusion is often accompanied by a reaction or sorption which 
can induce temperature inhomogeneities.  
Monte Carlo simulations 
of Lennard-Jones atoms in zeolite NaCaA are reported with
a hot zone presumed to be created by a reaction. 
Our simulations show that 
localised hot regions can alter both the kinetic 
and transport properties. 
Further, enhancement of the diffusion constant is greater for larger
barrier height, a surprising result of considerable significance
to many chemical and biological processes. 
We find an unanticipated coupling between reaction and diffusion 
due to the presence of hot zone in addition to that which normally exists
via concentration. 
\end{abstract}

\maketitle
      
Diffusion within porous materials or confined geometry is 
still poorly understood\cite{faraday,klafter}
despite increased attention in recent times\cite{bates,krishna}. 
Life sciences has a number of instances which relate to
diffusion within confined regions - for example, ion diffusion 
across  membranes and approach of a substrate towards an active site
of an enzyme \cite{lehnin}. 
Hydrocarbon separation and catalysis within zeolites  
provide instances of processes in chemistry \cite{karger}.
Problems involving fluid flow and excitonic transport
through porous medium are examples 
from physical sciences\cite{klafter}. 
The richness of the subject partly arises
from the  geometry of confined systems
( the fractal nature of the pores, for instance).
Further, while non-uniformity of concentration has
been dealt with in great detail, that of temperature has received 
little attention. 
In particular, when  
temperature is inhomogeneous, the very definition of diffusion as being an activated 
process needs a generalization. Such non-uniformity in temperature 
arises routinely in biological, chemical and  physical systems for variety of 
reasons. Here, we discuss issues relating to the 
possible sources of such hot spots and their influence 
on transport properties in the context of zeolites.

Zeolites are porous solids with pore sizes 
comparable to molecular dimensions\cite{barrer}.  
Due to its rich and diverse 
catalytic as well as  molecular sieve properties\cite{thomas1}
it has attracted much attention.  
The existence of specific catalytic and 
physisorption sites
coupled with their poor thermal conductivity could lead 
to local hot regions\cite{thomas1}.
(Typically in 10ps, the hot region decays less than a few percent).
This may affect both kinetic and diffusion properties. 
Such a situation can arise in many biosystems as well. For instance,
plasma membrane protein-encoding m-RNA IST2 
is shown to have high asymmetry in concentration between the 
mother cell and the bud\cite{ref6}. 
One possible way of maintaining such an asymmetry
against the concentration gradient is through 
localized hot or cold regions.
{\it In spite of the importance of such reaction induced hot spot 
and its influence on the diffusion of the species, this problem 
has not been addressed so far.}

Here, we study the effect of inhomogeneous 
temperature presumed to be created by  a `reaction', on the 
equilibration rate and self diffusion coefficient of guest molecules in 
zeolites. Monte Carlo
simulations on simple argon atoms in zeolite
A are reported here. Our results
show that  self diffusion coefficient $D$ is increased substantially due to 
the presence of a hot  zone. More significantly, {\it at a conceptual level} 
our analysis shows that local changes in temperature resulting from
reactions can induce additional coupling between reaction and diffusion.

\begin{figure}
\includegraphics{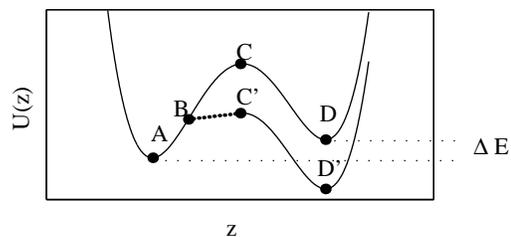}
\caption{\label{fig1}The effect of a hot zone at BC in the potential ABCD is to lower $D$ to $D'$.} 
\end{figure}

Landauer\cite{land75}, in a seminal paper, addressed the  
effect of a nonuniform
temperature bath on the relative occupation of competing local energy 
minima. For the case of a bistable potential $U(x)$ 
(the curve ABCD in Fig.~\ref{fig1}),  he showed that the presence of 
a localized heating in a region (say BC) lying between the lower energy
minimum A and the potential barrier maximum C can raise the relative
population of the higher energy minimum D over that given by
the Boltzmann factor $exp ( - \Delta\, E / k_B T )$. This has come to be known 
as the `blowtorch' effect \cite{land75}. Since this effect is rather counter 
intuitive, following Landauer, we convey the basic idea. Consider 
the motion of an overdamped particle in this potential 
(curve ABCD in Fig.~\ref{fig1}) subject to a uniform 
temperature $T_0$ along the coordinate. Then, the probability of 
finding a 
particle at $x$ is $P(x) \sim exp ( - U(x)/k_BT_0)$.  
Then the probability at A is higher than that at D. 
If the temperature in BC is raised to $T_b$,
then $P(x) \sim exp (- U(x)/k_BT_b)$ in $BC$, which is much smaller 
than $P(x)$ with $T=T_0$ and hence $log (P(x))$ is flat in BC.
Since  $-log P(x)$ can be considered as `potential' 
$U(x)/k_BT$, raising the temperature to $T_b$ 
in $BC$ is equivalent to modifying the `potential' to flatter 
curve $BC'$ (Fig.~\ref{fig1}). Since $P(x)$ is unaffected in other regions, 
the curve outside $BC$ will be the same excepting that the 
curve $CD$ would start at $C'$ and end at $D'$ such that 
$ U(x_C) - U(x_D) = U(x_{C'}) - U(x_{D'})$. Thus, the point $D$ 
is brought down relative to $A$. Consequently, $P(x_D)$ is higher than 
at the lower minimum $x_A$.

A decade later an appropriate diffusion equation for a nonuniform 
temperature profile 
was derived by van Kampen \cite{vkamp}, followed by additional work from 
Landauer himself\cite{land1}. Recently,  Bekele {\it et al.}\cite{Bekele} 
have shown that the escape rate is enhanced due to the presence of a hot zone. 
This implies that the rate of surmounting the barrier in zeolite will be 
increased by the presence of a hot zone and a consequent increase in 
diffusion constant. 

\begin{figure}
\includegraphics{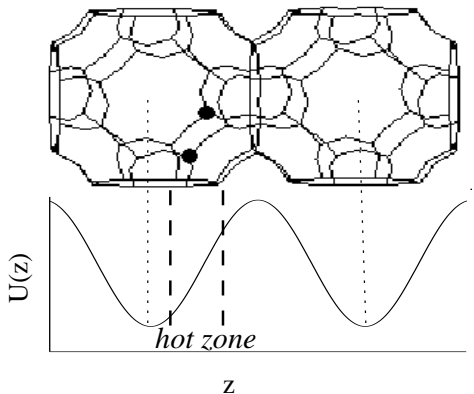}
\caption{\label{fig2} Two cages of zeolite NaCaA. $\bullet$ show reaction sites. The saddle point is at the window. A schematic one dimensional potential along the z-direction is shown below along with the induced hot zone. For set A simulations the potential at extremes of $z$ goes to infinity(not shown). 
For set B, the potential shown above is periodic.} 
\end{figure}

The physical system we simulate consists of NaCaA zeolite with
large ($\approx$ 11.5\AA\ dia) cages (the supercages) interconnected 
by shared narrow 8-ring windows ($\approx$ 4.5\AA\ dia). The potential
energy landscape has a maximum near the 8-ring window and a 
minimum located deep 
within the supercage.  A species arriving at a heterogeneous reaction site, 
assumed to be located between the window and the center of the cage, 
(Fig.~\ref{fig2}) releases a heat $q$ creating a local hot zone. 
Consequently, the molecule surmounts the barrier 
more easily.  Other molecules behind it also cross the barrier 
with relative ease due to the hot zone.
Here, we mimic the reaction by its principal effect - the presence
of a hot zone - by introducing it 
in between the potential maximum and minimum. 
In case of physisorption
 we assume that the heat released due to sorption is taken up by another
 diffusing particle. The effect arising out of the occupied volume of
 the adsorbent is ignored as it is negligible (for methane-NaY, this is
 typically 2\% of the cage volume).

We consider  Lennard-Jones particles confined to 
$ 2 \times 2 \times 1$ unit cells of NaCaA zeolite. 
There are 2$^3$ cages in each unit cell and
$4 \times 4 \times 2$ cages in the x, y and z-directions. 
The potential along the z-direction   
is a symmetric double well(Fig.~\ref{fig2}). 
Since the rate determining step is the passage through the 8-ring
window,  the distance of the particle from this plane
may be considered as the diffusion coordinate.
Due to lack of appropriate
techniques for including non-uniformity in temperature (that is,
in maintaining steady state excess temperature in a 
localised region) within
the existing molecular dynamics algorithms, we use Metropolis 
Monte Carlo algorithm in the canonical ensemble where the
total energy is 

\begin{equation}
 \Phi = \sum_g \sum_z \phi_{gz} + {1 \over 2} \sum_g \sum_g \phi_{gg} 
\end{equation}
\noindent
and\  
$\phi_{ab} = 4 \epsilon_{ab} [(\sigma_{ab}/r)^{12} - (\sigma_{ab}/r)^6];  a,b = g,h$
is the Lennard-Jones potential. Thus, the particle diffuses 
on the potential energy landscape of the zeolite.
The potential parameters for the guest-guest and guest-host interactions are 
$\sigma_{gg}$ = 2.73 \AA \  
 and  $\epsilon_{gg}$ = 0.9977 kJ/mol.
$\sigma_{O-O}$ = 2.5447 \AA,
 $\sigma_{Na-Na}$ = 3.3694 \AA, $\sigma_{Ca-Ca}$ = 3.35 \AA, 
$\epsilon_{O-O}$
 = $l$*1.28991 kJ/mol, $\epsilon_{Na-Na}$ = $l$*0.03924 kJ/mol 
and $\epsilon_{Ca-Ca}$ = $l$*9.5451 kJ/mol~\cite{du} where 
$l = 2,4$.  The cross terms are obtained from Lorentz-Berthelot 
combination rule. The cut-off distance was 12 \AA.

Two different 
sets of simulations A and B are carried out. 
In set A, for the calculation of the
escape rate, we impose 
periodic boundary conditions(PBC) along the x- and y-directions and 
a repulsive potential ($1/r^{12}$) 
at both ends along z-direction which
 enables a comparison with earlier 
work\cite{Bekele}. In set B, for the 
calculation of $D$ in three dimensions, 
PBC along the all the three directions is essential.
We investigate the influence of degree of hotness defined by 
$s = (T_b-T_0)/T_0$ and the barrier height $U_a$ ($\propto l\epsilon$)
on the equilibration rate and $D$. 
Here, $T_0$ and $T_b$ are the background and
blow torch temperatures respectively.  $T_0$ is kept at 300 K, and $T_b$
 is varied.

\begin{figure}
\includegraphics{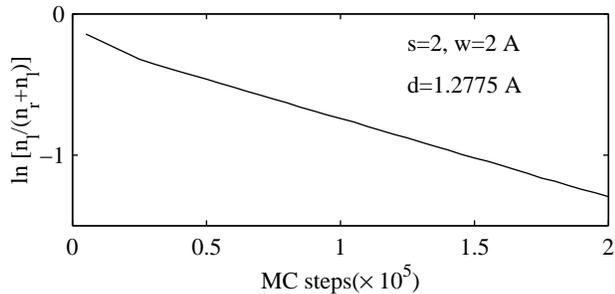}
\caption{\label{fig3} A typical plot of $ln (n_l/(n_l + n_r))$ \ vs. \ MC steps obtained from MC runs(set A).}
\end{figure}

Initially, all the 64 guest particles, at a concentration of 2 atoms/supercage 
are uniformly distributed in the  four left cages located along the $z$ 
direction (see Fig.~\ref{fig1}).  It is well known that transport properties 
are determined by the time scales associated with the approach to
the steady state. These are obtained 
by allowing the system to evolve towards  the steady state
in the presence of the hot zone.
Let $n_l$ and $n_r$ be respectively the total number
of particles in the cages to the left and the right of saddle 
point along the z-direction.
Then, the decay rate is obtained by
plotting the fraction of particles in 
the left cages as a function of time. A typical plot of 
$ln\ (n_l/(n_l+n_r)) \ vs.
\ t$ is shown in Fig.~\ref{fig3}.
The slope then gives the equilibration rate.  
The curve shown in Fig.~\ref{fig3} will reach a plateau at much longer
time scale.

\begin{figure}
\includegraphics{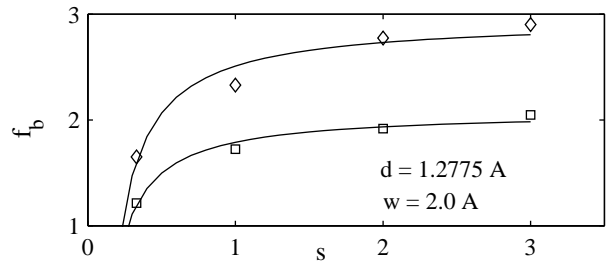}
\caption{\label{fig4}A plot of the enhancement factor, $f_b$, as a function of the degree of
hotness, $s = (T_b-T_0)/T_0$, for two different values of well depth, 
$l\epsilon$ ($l$ = 2($\square$), 4($\diamond$)) from MC runs(set A).}
\end{figure}

We consider the influence of the hot zone on the  
rate of approach to the steady state as a function of $s$
when the hot zone of width $w = 2$ \AA  \, is placed at a 
distance $d = 1.2775$ \AA \, from the window.
Instead of the escape rate, we use  
{\it the enhancement factor $f_b$}, which is the ratio of 
the escape rates from the left cage with and without the hot zone. 
A plot of $f_b$, as a function of $s$, for two 
values of $l$ is shown in Fig.~\ref{fig4}. 
The tendency to  approach saturation is seen for both 
values of $U_a$ ($\propto l \epsilon$), even though, it is less pronounced 
for the higher value ($l=4$). More importantly,
$f_b$ is a sensitive function of the barrier 
height\cite{Bekele}(Fig.~\ref{fig4}). Thus, the enhancement in
$f_b$ is greater when $U_a$ is larger which implies 
{\it larger enhancement in $D$ for systems 
with higher energy barriers}.

The significance of these results becomes apparent on examining a
real system such as methane in faujasite. 
The energy at the physisorption
site for methane in NaY zeolite (with Si/Al = 3.0) is -18 kJ/mol,
the energy difference between a free methane and a physisorbed methane. 
However, the energy released is significantly lower( $\sim$ 
-6 kJ/mol) when it is already within the zeolite\cite{nature}. This can
raise the temperature in the vicinity of the site.

We now consider the influence of a 
hot zone on diffusion through set B simulations for which
the starting configuration is the final
configuration of set A.
 We have studied the influence of 
$s$, and the barrier height. 
The ratio of the diffusion constant, $D_h$ 
with the hot spot to that without, $D_0$  
is enhanced in each case(Table~\ref{table1}).  
We note that
since all other conditions of the simulation are identical in the two
situations, $D_h/D_0$ is independent of the
basic length and time scales and also  
the details of the simulation such as the 
particle displacement(similarly for set A.)
Note that the {\em larger the  barrier height greater is the enhancement 
in $D$}, a result that has considerable implication.
These results are better understood on the basis of simple
arguments to estimate $D_h /D_0$. 

\begin{table}
\caption{\label{table1} $D_h/D_0$ for different sets of parameters. Here $d $ = 1.2775 \AA \ and $w$ = 2 \AA\ obtained from MC runs(set B).}
\begin{tabular}{|c|c|c|c|c|} \hline
$s$ &  $l\ (in$ & $n_r/n_l$ & $D_h/D_0$ & $D_h/D_0$ \\
 & $\epsilon = l*\epsilon_{gz})$  &  & (estimated) & (MC) \\ \hline
2 & 2 & 2.0175 & 1.5088 & 1.4283 \\
3 & 2 & 3.5416 & 2.2703 & 2.1971 \\
2 & 4 & 2.6132 & 1.8066 & 1.6747 \\ \hline
\end{tabular}
\end{table}

In the Kramers picture, the slowest time scale determining the 
approach to the steady state is identified with the escape rate. 
To estimate this, consider a
one dimensional symmetric double well potential with PBC. The 
rate equation for 
the number of particles to the left ($n_l$) and right ($n_r$) of
the potential maximum, is 
$\dot n_l = - w_{lr}n_l + w_{rl}n_r,$ where $w_{lr}$ and $w_{rl}$
are the escape rates from the left to the right well 
and vice versa respectively.
These are also the very time scales that determine 
the steady state through detailed balance condition:   
$\dot n_l = 0 $ or $w_{lr}n_l = w_{rl}n_r$. 
In the absence of hot spot,
we have $w_{lr} = w_{rl}$. Then the mean escape rate from a well is
 $w_0 = (w_{lr} + w_{rl})/2 $.
Then the diffusion constant
in one dimension, $D_0 = a^2 w_0/2$, 
 where $a$ is the distance between the two minima.
On introduction of a hot zone in the left well, $w_{lr}$
has been shown to increase considerably, while $w_{rl}$
increases only marginally \cite{Bekele}, and to a good approximation
$w_{rl} \approx w_0$.  
The mean escape rate is $w_h = (w_{lr} + w_{rl})/2$. 
Using this with $w_{lr} = w_{rl} n_r/n_l$, 
we get $D_h = a^2 w_h/2 \approx D_0 ( 1 + n_r/n_l)/2$. 
Since, $n_r/n_l > 1$ in the presence of a hot zone, it is evident
that $D$ is enhanced. 
Note that  this refers to a nonequilibrium inhomogeneous
situation as $n_r/n_l$ refers to the steady state which can only be 
obtained numerically.
Table~\ref{table1}, shows the values of $D_h/D_0$ obtained from the mean square 
displacement through MC simulations for three sets of parameter values 
along with  $D_h/D_0$  estimated  from the above expression using the  
steady state values of $n_r/n_l$ from MC simulations. (Note $n_r$ and $n_l$ 
refers to the total number of particles to the left and right of 
the potential maxima between the two cages in the z-direction.)
Note that the estimated values are 
close to MC values.

To facilitate comparison of these results with real systems, 
we have estimated the likely 
increase in temperature when hydrocarbons and 
other guest species are sorbed within zeolites such as NaX. We have listed 
in Table~\ref{table2} isosteric heat of sorption ($\Delta H_{ads}$) of some linear 
alkanes, Xe and water within faujasites. We have also listed the mean 
heat capacities ($C_m$) of the guest-zeolite systems \cite{barrer}. 
From these data, the maximum increase in $\Delta T$ can be estimated from
$ \Delta T = T_b - T_0 = \Delta H_{ads}/C_m $ which
is in the range 820 K to 2028 K (Table~\ref{table2}). 
Thus, the parameter $s$ varies from 1.7 to 6.7 
for which $f_b$ 
can be as large as  3 for $s  \sim6$ and $l = 2$. 
Since $f_b$ is determined by 
the very rate constants that lead to {\it steady state}, 
it also implies that the 
diffusion constant in the inhomogeneous medium can increase by a
factor of two, 
even for moderate values of $s$ as can be seen from Table~\ref{table1}.

\begin{table}
\caption{\label{table2} Expected rise in temperature and $s $ for typical guests
when adsorbed in common zeolites estimated from heat of adsorption,
$\Delta H_{ads}$ and the mean heat capacity, $C_m$ data.}
\begin{tabular}{|c|c|c|c|c|c|c|}   \hline
\multicolumn{2}{|c|}{System}& $\Delta H_{vap}$ & $\Delta H_{ads}^a$ & $C_m$ & $T
_b-T_0$ & $s$ \\ \cline{1-2}
guest&zeolite&kJ/mol&kJ/mol&J/mol.K&K& \\ \hline
n-C$_4$H$_{10}$ & Na-X & 66 & 174 & 105 & 1689$^b$ & 5.6 \\
n-C$_7$H$_{16}$ & Na-X & 87 & 228 & 176 & 1809$^b$ & 6.0 \\
n-C$_7$H$_{16}$ & Na-X & 87 & 228 & 209 & 1090$^c$ & 3.3 \\
neo-C$_5$H$_{12}$ & Na-X & 54 & 130 & 129 & 1011$^b$ & 3.3 \\
iso-C$_8$H$_{18}$ & Na-X & 88 & 246 & 185 & 1329$^d$ & 4.0 \\
Xe & Na-Y & --  & 18 & 22 & 820$^e$ & 1.7 \\
H$_2$O & Na-X & --  & 142 & 70 & 2028$^b$ & 6.7 \\ \hline
\end{tabular}

\noindent
$^a$Calculated from $\Delta H_{vap}$ and the ratio of $\Delta H_{ads}$ to $\Delta H_{vap}$\cite{barrer}.
$^b$ $T_0$=300K; $^c$ $T_0$=333K; $^d$ $T_0$=325K; $^e$ $T_0$=473K.

\end{table}

An interplay of reaction and diffusion is known to give rise to complex 
dynamics which can manifest in different ways\cite{pearson1}.
The product profile in a reaction 
is controlled by the diffusion rate of the product species formed
rather than the reaction rate as is the case with the formation of 
p-xylene 
in ZSM-5\cite{thomas1}. Due to low diffusivity of 
ortho and meta isomers, they are not observed as products
even though they are formed.
The coupling  
between reaction and diffusion in such systems is
via the concentration of 
the reactant and product species \cite{demo}. In contrast, 
in the present situation, the enhancement in $D$ is a direct
consequence of inhomogeneous temperature.
This study 
demonstrates that 
such a coupling between reaction and diffusion can arise not just due 
to concentration, but also due to the increase in local 
temperature, {\it a fact that could not have been anticipated}. 
These results also show how diffusion is increased 
in the presence of physisorption or chemisorption which are usually 
exothermic.  $D$ may decrease if the reaction is endothermic.

The present analysis can provide an insight into a well known
experimental observation where a warm adsorption front is seen
to move rapidly during the adsorption of a gas into an evacuated
zeolite\cite{basma}. As the initial molecules arrive at a 
physisorption site, heat is released which 
aids the molecules at the front
to cross over the energy barrier and propel the gas forward.
Zeolites are crystalline solids, and hence the active sites
are located in a periodic manner. Thus, as the front moves further
into the zeolite, hot zones are created successively providing
a periodic driving force for diffusion which
is in addition to that arising from the concentration gradient. 
This explains the rapid movement
of the front.

Thus, {\it influence of a reaction induced hot zone 
can affect diffusion in ways which will
be important both from fundamental as well as industrial 
perspective.}  Though our results are obtained in the context of zeolites, 
it is evident that they are of significance to many biological
processes where concentration gradients
are frequently accompanied by difference in temperature. 
These results also have implication to the separation of 
isotopes\cite{yamakawa},
the petrochemical industries, fast ion conducting battery materials, etc.
We believe that in the foreseeable future,  local inhomogeneities in
temperature will be exploited in a number of ways to bring forth
novel processes. 

\noindent {\it Acknowledgments }  : AVAK \& SY thank
IUC, Indore for the award of Project Assistanship and 
financial support from DST, 
New Delhi ( project no.  SP/S1/H-12/99).

\end{document}